\documentclass[acmsmall,authorversion]{acmart}
\usepackage{amsmath}
\usepackage{graphics}
\usepackage{caption}
\usepackage{subcaption}
\usepackage{relsize}
\usepackage{sparklines}
\usepackage[inline]{enumitem}
\newlist{inlinelist}{enumerate*}{1}
\setlist[inlinelist]{label=(\roman*)}

\setcopyright{acmlicensed}
\acmJournal{TORS}
\acmYear{2023}
\acmVolume{1}
\acmNumber{1}
\acmMonth{8}
\acmPrice{15.00}
\acmDOI{10.1145/3613455}

\newcommand{\ExpSym}[1]{\operatorname{E}}
\newcommand{\Exp}[2][]{\ExpSym{}\if\relax\detokenize{#1}\relax\else_{#1}\fi\left[#2\right]}
\newcommand{\CondExp}[3][]{\Exp[#1]{#2 | #3}}

\newcommand{\UtilSym}{M}

\newcommand{\System}{S}
\newcommand{\User}{u}

\begin{document}

\title{Distributionally-Informed Recommender System Evaluation}

\author{Michael D. Ekstrand}
\email{ekstrand@acm.org}
\affiliation{%
  \institution{People \& Information Research Team, Boise State University}
  \city{Boise}
  \state{ID}
  \country{USA}
  \postcode{83705}
}
\affiliation{%
  \institution{Dept. of Information Science, Drexel University}
  \city{Philadelphia}
  \state{PA}
  \country{USA}
  \postcode{19104}
}

\author{Ben Carterette}
\email{carteret@acm.org}
\affiliation{%
  \institution{Spotify}
  \country{USA}
}
\author{Fernando Diaz}
\email{diazf@acm.org}
\affiliation{%
  \institution{Language Technologies Institute, Carnegie Mellon University}
  \city{Pittsburgh}
  \state{PA}
  \country{USA}
}


\begin{abstract}
  Current practice for evaluating recommender systems typically focuses on point estimates of user-oriented effectiveness metrics or business metrics, sometimes combined with additional metrics for considerations such as diversity and novelty.
  In this paper, we argue for the need for researchers and practitioners to attend more closely to various \textit{distributions} that arise from a recommender system (or other information access system) and the sources of uncertainty that lead to these distributions.
  One immediate implication of our argument is that both researchers and practitioners  must report and examine more thoroughly the distribution of utility between and within different stakeholder groups.
  However, distributions of various forms arise in many more aspects of the recommender systems experimental process, and distributional thinking has substantial ramifications for how we design, evaluate, and present recommender systems evaluation and research results.
  Leveraging and emphasizing distributions in the evaluation of recommender systems is a necessary step to ensure that the systems provide appropriate and equitably-distributed benefit to the people they affect.
\end{abstract}





\maketitle

\section{Introduction}


Recommender systems and related information access systems, such as search engines, are large research areas and massive industries.
They are the backbone of many of the services we now use daily, from news to music recommendations.
As such, they have an indelible effect on the lives of both consumers (users) and producers.
The processes by which we decide how to deploy and use these systems impact consumers and producers, potentially in major ways.
In order to understand this impact, we first have to be able to evaluate the systems.

Evaluation of recommender systems as practiced today has roots in the Cranfield experiments to evaluate search systems done by Cyril Cleverdon and colleagues in the 1960s \citep{cleverdonCranfieldTestsIndex1967}, as well as supervised machine learning evaluation.
Cleverdon et al. evaluated ``indexing devices'' by their ability to improve precision and recall of relevant research papers in keyword searches.
This practice evolved in the 1970s with Salton's experiments on SMART \citep{saltonSmartProjectAutomatic1991}, and further evolved with the introduction of standardized test collections, catalogues of evaluation measures, and statistical significance testing, reaching a culmination in the 1990s with TREC.
TREC introduced a fully standardized evaluation methodology for search that is now widely-used in recommender systems research (as summarized by \citet{herlockerEvaluatingCollaborativeFiltering2004} and \citet{gunawardanaEvaluatingRecommenderSystems2022}) in addition to information retrieval work. This methodology has been adopted in commercial industry for offline evaluation and further explored in contexts such as its ability to predict user or expert evaluation results \citep[e.g. ][]{koukiLabProductionCase2020}.
Standard evaluations essentially compute a pointwise estimate of one or more evaluation metrics.
These metrics are typically focused on the experience of one class of stakeholders, and decisions about the relative usefulness of systems is made on the basis of comparing these estimates.


In this paper we argue that pointwise effectiveness estimates are not sufficient for either reporting research results or for making decisions in production environments.
Our proposal is that recommender system and search evaluation should, indeed, strive for a different target: it should attend to the distributions of these metrics to understand how the system impacts different users, producers, and other stakeholders, and make deployment decisions in light of a holistic consideration of the effects of proposed technologies across the individuals and organizations participating in an ecosystem.

Our perspective is that thinking only in averages is harmful to recommender system research and applications.
\citet{fuhrCommonMistakesIR2017} listed some problems with search evaluation, including over-precise results and a lack of reporting effect sizes (and Sakai's response \citep{sakaiFuhrGuidelineIR2020} agrees with some and disagrees with others), many of which also apply to recommendation; we believe many of these problems and disagreements would likely vanish if we as research and practitioner communities agreed on the use of distributions rather than averages in evaluation, reporting, and decision-making.


\section{Current Practice and Limitations}
\label{sec:current}

The current standard evaluation methodology is this: given a system $S$, an evaluation measure $M$, a set of relevance signals $R$, and a set of requests $Q$ (each consisting of a user with their past history, possibly accompanied by context and/or implicit or explicit data about session intent, such as a query or initial interactions), collect the output of $S$ for each $q \in Q$ --- let us call it $S_q$ --- and compute $M(S_q, R_q)$.  
The effectiveness of $S$ is then estimated by the mean of $M(S_q, R_q)$ over all $q \in Q$.  
We refer to this as a pointwise estimator, denoted by $\widehat{M_S}$.

Pointwise estimators are useful because they enable researchers and practitioners to perform unambiguous comparisons between systems. A group of systems can be ordered by this pointwise estimator, ``winners'' can be declared, straightforward decisions can be made about which systems to deploy to users, and so on.
Using the mean for the pointwise estimator is particularly useful because the sample mean, as a statistic, has certain desirable properties --- it reflects the central tendency of the measurement, it tends to a normal distribution in the limit (when distributions of measurements are well-behaved), and it is sufficient (in that no other statistic is necessary) to estimate the central population tendency.

We can further compute other statistics of effectiveness, such as the standard deviation, and use them in statistical significance tests like the $t$-test if we would like analysis or decisions to be a bit more robust; reporting with confidence intervals can provide further information on the precision of these estimates.
Online evaluation is not really different except that relevance signals are more directly positive user signals such as clicks or purchases.
Multiple metrics may be employed, often in a multi-objective framework \citep[e.g.][]{vandoornBalancingRelevanceCriteria2016,ribeiroParetoefficientHybridizationMultiobjective2012}, but the focus is usually on individual points in the evaluation metric space.

Despite its simplicity and power, the approach of comparing systems and making decisions using means alone (or in conjunction with outcomes of statistical significance tests) has some limitations:
\begin{itemize}
\item It only considers one perspective, that of the user interacting with the results.  Different metrics may model these interactions in different ways, but regardless they ignore the perspective of producers and other stakeholders.
\item Generally speaking, it only considers one metric.  Though other metrics may be part of a larger argument or decision process, there is generally not a principled approach to comparing multiple metrics.
\item It treats users as interchangeable by abstracting the user experience into a model of interaction with system results.
\item It treats all the components of the experimental environment and system outputs as deterministic and certain when there may in reality be uncertainty, vagueness, ambiguity, arbitrariness, and randomness at many points in an experiment.
\item It collapses the varied experiences of different stakeholders into a single measurement measurement. For example, it applies a metric based on a single model of user interaction uniformly across all users and system results, aggregating into a point estimate, in a way that obscures how the system may impact different users (or providers) differently.
\item It collapses time into one snapshot by either taking a single day measurement or averaging over a period of time.
\end{itemize}

Problems compound when the assumption is made that improving effectiveness by some metric on average improves the value to users. 
There are many reasons why this may not be so, not least of which is that a pointwise average effectiveness may not map to any individual users' experience of the system --- there is no such thing as an ``average user''!
Any change is likely to impact some users positively and some negatively, and even a statistically significantly positive change may present risks to some of the users --- to say nothing of other stakeholders.
Despite this, there is currently no widespread effort to more deeply understand search and recommender system effectiveness.

The simplicity and power of the mean combined with the hidden or unstated assumptions we detail above could be seen as enabling a scientific culture of ``leaderboard chasing'' or ``state-of-the-art (SOTA) chasing''.
Since it is very easy to compare means over a standard test set and declare a winner, it follows that it is easy to optimize for the mean without ever understanding the data, the setting, or the potential users of the system.
Several authors have independently argued both against the culture of SOTA chasing~\cite{adomaviciusStabilityRecommendationAlgorithms2012,larsonAnotherChecklistNew2022,churchEmergingTrendsSOTAChasing2022,rodriguezEvaluationExamplesAre2021} and for the use of alternative evaluation frameworks based on deeper analysis.  In particular, \citet{rodriguezEvaluationExamplesAre2021} describe an evaluation framework called {\em DAD}, for Difficulty and Ability Discriminating leaderboards, and \citet{jannachMoreImpactfulRecommender2019} argue for evaluating scientific work by {\em impact}, which includes measurement but also value, risk, methods, and more.  Our contribution is not an evaluation framework, but an argument for making greater use of raw distributions and a greater variety of distribution statistics and visualizations to analyze and understand the effectiveness of a recommender system.

Accounting for uncertainty is one important aspect of moving beyond simplistic comparison of means.
For example, the \textit{rank-biased precision} (RBP) measure of \citet{rbp} is characterized by a user model of behavior that includes a random chance of abandoning the ranking at any point.  Similar measures (ERR~\cite{err}, EBU~\cite{ebu}) incorporate more complex probabilistic user models.  However, the final effectiveness measures themselves are still computed as pointwise expectations.
One notable exception is provided by \citet{wangRobustRankingModels2012}, who proposed ``helped-hurt histograms'' that show the distribution of change in performance over users or queries.

Measures of result diversity often include a probability distribution over different possible query intents, along with relevance judgments to those intents---the $\alpha$-nDCG measure~\cite{alpha-ndcg} is the classic example, with measures like ERR-IA~\cite{err-ia} following suit. 
Again, these measures are in practice computed as expectations over the intent distributions, discarding any distributional information in the final reporting.

Distributional information is also used in statistical significance tests, where it is a component of computing a $p$-value.
In reporting results, however, the distributions are discarded in favor of the $p$-value or a simple indicator of statistical significance.
Bayesian evaluation that reports posterior distributions does exist \citep{ekstrandExploringAuthorGender2021}, but is rare.

Collectively, these observations suggest three classes of distributions we should consider: 
\begin{inlinelist}
    \item \textbf{sample distributions} that capture uncertainty obscured by point estimates, 
    \item \textbf{sub-group distributions} that capture sub-group performance obscured by aggregation, and
    \item \textbf{stakeholder distributions} that capture stakeholder performance obscured by omission.   
\end{inlinelist}
While individually touched on by prior work, these classes have not been treated as an evaluative paradigm acknowledging that a more granular description of systems on their impacts.  
By analyzing and reporting on the uncertainty, we achieve greater transparency, better scientific practice, and create new opportunities for research and development in recommender systems and related research.




\section{A Vision for Thorough Evaluation}

As described in Section~\ref{sec:current}, although the most common paradigm for recommender systems is to report the mean of one or more performance metrics, averaged over test instances (e.g. users), some work has addressed some classes of our concerns. For example, more rigorous evaluation reports the results of a statistical analysis of mean performance, such as a significance test or a confidence interval \citep{carteretteStatisticalSignificanceTesting2019} (although \citet{ihemelanduStatisticalInferenceMissing2021} observe that this is often overlooked in the published research literature). Other work includes ablation studies, where the impact of individual components on this performance metric yields insight into their various contributions \citep{mehrotraAuditingSearchEngines2017,ekstrandAllCoolKids2018,ekstrandExploringAuthorGender2021,ferraroBreakLoopGender2021}.  Recent work in multi-stakeholder recommendation seeks to broaden our understanding of who is impacted by systems \citep{abdollahpouriRecommenderSystemsMultistakeholder2017}.

While these isolated methods are steps toward address these classes of uncertainty, and \citet{tagliabueEvalRSRoundedEvaluation2022} integrate some of these ideas into a multi-faceted evaluation, we envision the possibility of comprehensive evaluation reports that describe a wide range of aspects of the performance and behavior of a recommender system (or other information access system, such as a search engine or information filter), that provide future researchers and practitioners with knowledge that enables them to more carefully assess the applicability of a proposed development to their context, and to understand the behavior of a potential system in the context of a wide range of business and social goals. This flows from distributional analysis: reporting and attending to the distribution of system performance and behavior metrics over a range of axes, through both reporting of distributions themselves (in distribution plots and computationally-useful representations) and richer sets of statistics describing these distributions. Such evaluations will allow for many current and new questions to be answered, including:

\begin{itemize}
\item How is system performance distributed among users, information needs, and/or items? Does it perform relatively well for most users, or are some use contexts left behind?
\item Does it perform comparably well across groups of users, item producers, or other stakeholders, or does the short end of variation in performance systematically fall on groups that are often also marginalized in society?
\item When comparing two systems, how is the improvement distributed? Does it benefit many people, or provide substantial improvement for a few while reducing utility for others?
\item How confident can we be in the apparent improvement? Is it robust over a range of assumptions and likely to be replicable?
\item How dependent is the reported performance on the uncertainties associated with missing data, erroneous data, and other sources of bias and uncertainty in the system’s training and evaluation data?
\item How stable are the reported performance results under data resampling, re-training with different random seeds, and other sources of variability?
\end{itemize}

We do not claim that this will make evaluation easier; in fact, the increased richness of reporting experimental results will require subtlety and care to properly interpret with respect to particular goals and tasks.
However, it will enable the community to make a more thorough accounting of system behavior and performance, enabling richer follow-on analysis and more robust matching of systems to application requirements.

\section{Sources of Uncertainty}
\label{sec:sources}

Our central contention is that recommender system evaluation needs to look beyond such pointwise estimates of individual metrics, possibly combined with statistical measurements of confidence or precision, and consider more fully \textit{distributions} of performance.
These distributions, broadly speaking, characterize uncertainty about the results: we do not know, precisely, how well a system will perform in aggregate, or how well it will perform for either a fixed or random user.

Uncertainty comes in various forms, which can be broadly categorized \citep{hullermeierAleatoricEpistemicUncertainty2021} into \textit{epistemic} uncertainty, where we lack knowledge about an aspect of the data, information need, etc.; and \textit{aleatoric} uncertainty, where there is a random aspect of the system and its context of use that is either intrinsically random (and therefore unmodelable even with perfect knowledge) or would require modeling outside the reasonable scope of the system.\footnote{\citet{hullermeierAleatoricEpistemicUncertainty2021} define aleatoric uncertainty only as \emph{intrinsically} random such that perfect knowledge cannot remove the uncertainty, but this opens many philosophical questions about the nature and existence of randomness.  For our present purposes, however, these questions are not relevant, and it suffices to consider external factors that a reasonably complete information access system would not attempt to computationally model as aleatoric, even if advanced knowledge of natural or human phenomena may theoretically make them modelable.}
For the present purposes, we consider most kinds of variance, such as variance between users or topics (e.g. varying topic difficulty), to be aleatoric uncertainty by assuming the arrival of users or queries to be an inherently random process; grouping it in this way vs. treating variance as a third source of ``uncertainty'' producing distributions does not alter our core argument.
Any of the forms of uncertainty we discuss can be analyzed at the level of sample distributions; many also admit subgroup distributions, and some of them admit stakeholder distributions.

In this section we describe sources of uncertainty throughout the recommender system deployment and evaluation processes: what aspects of a system result in a distribution of utility or performance?

\subsection{Experimental Process}
The first source of distributions comes from randomness in the experimental process: when a data set is randomly split into train, validation, and test subsets, different splits may produce different effectiveness results, both due to training the model on a different set of data (so its output may differ) and testing on a different set of test requests.
This variance in retraining over different training samples is the source of variance discussed in the bias-variance tradeoff and is a source of aleatoric uncertainty.
There may also be variance as when repeatedly training and evaluating the same model on the same data set with different random seeds affecting initial conditions, stochastic training order, etc. \citep{antoniakEvaluatingStabilityEmbeddingbased2018}.
Some models will also produce different results with different training data and random seeds.

There is also epistemic uncertainty around the correctness or appropriateness of different experimental decisions, such as data splitting strategies or metric parameters.
Modeling this uncertainty, and running experiments with multiple settings, can enable decisions that account for the uncertainty in evaluation design.

\subsection{Users, Contexts, and Intents}
Users, along with their behavior, preferences, and the contexts and intents with which they use the system provide several additional sources of uncertainty.
In production, a system will respond to requests (users seeking information, possibly with explicit queries and/or contextual variables to further inform the system of their specific information need) as they arrive, and the precise sequence of requests is a form of aleatoric uncertainty we refer to as \textit{request uncertainty}.

In a typical evaluation, such as a top-$N$ recommender evaluation or a TREC-style IR evaluation, the system produces a ranked list of results for each request in the test data, and its effectiveness is measured with a metric like nDCG or MRR.
This set of test requests is often treated implicitly as a random sample from the population of possible requests \citep{smuckerComparisonStatisticalSignificance2007}.
System effectiveness may vary widely from need to need; the nature, shape, and effects of this distribution are often lost in a pointwise aggregate.
Two systems with the same mean nDCG may have very different distributions of that utility, which results in significantly different experiences for users (or users with different queries or contexts), even though expected utility (as captured by nDCG) is equal; we show an example of this in Section~\ref{sec:tools:inspection}. 


Once the system has received a particular request, that request is still incomplete and carries a tremendous amount of uncertainty.   Requests, especially coarse representations of preference or context or discrete query strings, can collapse multiple user intents and, as a result, introduce uncertainty about which items are relevant and which are not.  We refer to this as \textit{target uncertainty}.  TREC initiatives use the practice of determining relevance based on whether a document contains \textit{any} relevant material.  Guidelines for web search relevance labels encode intent distributions into item ratings, with higher grades reflecting popularity of that intent \cite{googleSearchQualityEvaluator2022}.  These methods for dealing with ambiguity collapse a distribution of performance across intents into scalar numbers. 

In offline evaluation, user browsing models are the foundation of most metrics \cite{sakai:user-models,carteretteSystemEffectivenessUser2011a,carteretteIncorporatingVariabilityUser2012}.
Simple position discounts reflect a distribution of stopping behavior.  Although often considered measures of utility, this perspectives allows us to interpret metrics as point estimates over user behavior.  We refer to this as \textit{behavioral uncertainty}, and it is typically epistemic.  Even though most salient in offline metrics \citep{breeseEmpiricalAnalysisPredictive1998}, this can also be encoded in the assumptions, weights, and formulae in online evaluation \cite{chapelleDynamicBayesianNetwork2009}. 

The labeling process itself --- conducted by raters in offline evaluation or derived from behavior in online evaluation --- can also carry uncertainty.  There may be inconsistency across raters in assessing relevance for a request \cite{carteretteHereThere2008}.  Behavioral data such as clicks and streams are inherently noisy.  We refer to this epistemic uncertainty as \textit{label uncertainty}.  While label uncertainty is seldom modeled explicitly, it can be quantified in a Bayesian paradigm with distributions over the relevance of an item to a need \citep{carteretteBayesianInferenceInformation2015}; \citet{huCollaborativeFilteringImplicit2008} use a simple approximation of label uncertainty that interprets positive observations through the lens of ``confidence'' in their implicit-feedback collaborative filter (observed items have a high confidence of relevance, and unobserved items have a low but nonzero confidence).

So far, we have discussed uncertainty in evaluating an individual request (a sample distribution).  We can also consider uncertainty when evaluating systems over a population of requests, perhaps from multiple users. 

To start, requests are not independent and arise from often-unobserved structure, obfuscated in point estimates.  Requests can be structured or sliced from a variety of perspectives, depending on the goal of the analysis; this yields subgroup distributions.
Users, whether they manifest as collections of requests (as in information retrieval) or individual requests (as in common recommendation paradigms), can be grouped along multiple different and intersectional dimensions, dictated by a social or demographic perspective of interest.  We refer to this as \textit{user group uncertainty}.  The distribution of utility across this structure can surface systematic differential performance.  For example, \citet{mehrotraAuditingSearchEngines2017} studied the distribution of search engine quality across demographic groups, and \citet{ekstrandAllCoolKids2018} did the same for top-$N$ recommendation.

In a search context, queries can be grouped by session or task \cite{jones:session-time-outs}, which can then be grouped by individual user.  We refer to this as \textit{individual user uncertainty}.  Requests can also be grouped by request type  \cite{broderTaxonomyWebSearch2002} or the semantics of the information need (e.g. topic or product category).  We refer to this as \textit{request group uncertainty}.   

For each of these types of analyses, we are effectively computing the distribution of utility conditioned on a particular variable, with the mean representing the conditional expectation (e.g. aggregating by user gives us $\CondExp[\System]{\UtilSym}{\User}$), and we can then examine the conditional distribution of that measurement over the set of users (or queries, sessions, etc.).
In this way, distributional analysis is a vital tool for capturing the way the system's impact, such as utility with respect to the user's information need, is distributed across the system's various users, and identifying groups of users who are left out or under-served.

\subsection{Items}
Items may also bring uncertainty in various ways.
For one way, the set of items may be a sample from a larger population, bringing aleatoric uncertainty when the experiment or system is re-run on a different sample.

There may also be epistemic uncertainty in understanding the items themselves.
While the item's content (e.g. a document's text or a video's audiovisual content) is often certain, user-contributed data, such as tags and categories (``folksonomies'' \citep{xuExploringFolksonomyPersonalized2008,petersFolksonomyInformationRetrieval2008}), may result in uncertainty about item attributes; such attributes may also be uncertain even when provided by trained experts.
We call this \emph{item feature uncertainty}.
This uncertainty can also arise from inference techniques for items, such as object recognition in visual items (with the line between this and item-oriented model uncertainty in the next section admittedly blurry).

Further, as with users, we can also compute distributions of item-side effects such as exposure \citep{diazEvaluatingStochasticRankings2020} over the various items or item providers (such as recording artists, film producers, or authors) and their attributes.
This forms the basis of understanding how the benefits the system provides to the people who create and produce the items it recommends are distributed across those people both individually and with respect to socially-salient group identities \citep{fnt-fairness,rajMeasuringFairnessRanked2022}.

\subsection{Algorithm}
Information access algorithms themselves can additionally introduce (and, in some cases, account for) uncertainty.
Various aspects of a recommendation model may have epistemic uncertainty in their internal representations and/or outputs.  This can apply to any modeling component in the system, including query intent models, user models, context models, item models, and relevance models.
We refer to this as \emph{model uncertainty}.
This uncertainty can arise from uncertainty in the data that propagates through to the model, or uncertainty that arises through the model's attempts to interpret ambiguous or contradictory signals.

In some situations, the algorithm is designed to be random.  
We refer to this as \emph{stochastic algorithm uncertainty}.
Stochasticity can be useful for a variety of reasons, including diversity \citep{lathiaTemporalDiversityRecommender2010}, exploration of policy spaces \citep{radlinskiLearningDiverseRankings2008}, and  to more equitably distribute subtractable goods \citep{ostromRulesGamesCommonPool1994} such as recommendation opportunities among competing content providers \citep{diazEvaluatingStochasticRankings2020}.

Model uncertainty and stochastic algorithm uncertainty give rise to sample distributions, where the samples are either runs of an experiment or draws from the model's stochastic distribution.


\subsection{Simulations}
\label{sec:simulation}
Lastly, some experimental designs use probabilistic simulations that introduce further uncertainty in their results.
There are a range of types of simulation \citep{ekstrandMultiversalSimulacraUnderstanding2021}, such that any offline evaluation can be characaterized as a kind of simulation \citep[\S 2.5]{fnt-fairness}; others run a traditional evaluation repeatedly over synthetic data, simulate an entire information access feedback loop, or simulate experimental outcomes.
Simulation has proven a valuable tool for studying the behavior of statistical techniques \citep{urbanoStatisticalSignificanceTesting2019,paraparUsingScoreDistributions2020} and the effects of missing data on evaluation outcomes \citep{tianEstimatingErrorBias2020}, among other experiments.

These simulations introduce both aleatoric uncertainty through their use of random data (different runs will have different outputs; we call this \textit{stochastic simulation uncertainty}), and epistemic uncertainty about the data generating process and particular parameter settings that best match the simulation to the world and provide external validity for its results (\textit{simulation parameter uncertainty}).
Tuning the simulation based on system logs \citep{mcinerneyAccordionTrainableSimulator2021} and optimizing parameters to produce data that mimics existing data sets \citep{tianEstimatingErrorBias2020} can reduce but not eliminate this epistemic uncertainty.





\section{Tools for Distributional Evaluation}



Considering distributions in recommender system evaluation requires expanding our toolbox for analyzing and reporting the results of our evaluations.  This applies both for internal analyses and reports to evaluate systems for production use, and for publications in venues such as ToRS, RecSys, and SIGIR.  Some tools are readily available, at least in a basic form, while others may require further research to develop best practices to give readers and decision-makers a more comprehensive view of system behavior.
Our case study in Section~\ref{sec:study} demonstrates some of the available tools more thoroughly.

\subsection{Graphical Inspection}
\label{sec:tools:inspection}

\begin{table}[tbph]
    \centering
    \begin{tabular}{lccccc}
Algorithm
& Mean
& 10\%ile
& Median
& 90\%ile
& Dist. (KDE) \\
\toprule
IALS &
\begin{tabular}{@{}c@{}}%
0.061 \\%
{\smaller[2] (0.057, 0.065)}\end{tabular} &
\begin{tabular}{@{}c@{}}%
0.000 \\%
{\smaller[2] (0.000, 0.000)}\end{tabular} &
\begin{tabular}{@{}c@{}}%
0.021 \\%
{\smaller[2] (0.017, 0.027)}\end{tabular} &
\begin{tabular}{@{}c@{}}%
0.173 \\%
{\smaller[2] (0.166, 0.185)}\end{tabular} &
\begin{sparkline}{9}
\sparkdot 0.061 0.160 red
\spark
  0.000 0.575
  0.020 0.469
  0.041 0.234
  0.061 0.160
  0.082 0.143
  0.102 0.126
  0.122 0.115
  0.143 0.109
  0.163 0.103
  0.184 0.076
  0.204 0.050
  0.224 0.039
  0.245 0.032
  0.265 0.027
  0.286 0.022
  0.306 0.015
  0.327 0.010
  0.347 0.008
  0.367 0.003
  0.388 0.001
  0.408 0.000
  0.429 0.001
  0.449 0.001
  0.469 0.000
  0.490 0.000
  0.510 0.000
  0.531 0.000
  0.551 0.000
  0.571 0.000
  0.592 0.000
  0.612 0.000
  0.633 0.000
  0.653 0.000
  0.673 0.000
  0.694 0.000
  0.714 0.000
  0.735 0.000
  0.755 0.000
  0.776 0.000
  0.796 0.000
  0.816 0.000
  0.837 0.000
  0.857 0.000
  0.878 0.000
  0.898 0.000
  0.918 0.000
  0.939 0.000
  0.959 0.000
  0.980 0.000
  1.000 0.000
  /
\end{sparkline}
\\
IKNN &
\begin{tabular}{@{}c@{}}%
0.057 \\%
{\smaller[2] (0.053, 0.061)}\end{tabular} &
\begin{tabular}{@{}c@{}}%
0.000 \\%
{\smaller[2] (0.000, 0.000)}\end{tabular} &
\begin{tabular}{@{}c@{}}%
0.012 \\%
{\smaller[2] (0.009, 0.017)}\end{tabular} &
\begin{tabular}{@{}c@{}}%
0.163 \\%
{\smaller[2] (0.160, 0.170)}\end{tabular} &
\begin{sparkline}{9}
\sparkdot 0.057 0.138 red
\spark
  0.000 0.634
  0.020 0.474
  0.041 0.209
  0.061 0.130
  0.082 0.109
  0.102 0.103
  0.122 0.103
  0.143 0.120
  0.163 0.130
  0.184 0.081
  0.204 0.041
  0.224 0.028
  0.245 0.023
  0.265 0.023
  0.286 0.022
  0.306 0.015
  0.327 0.007
  0.347 0.005
  0.367 0.005
  0.388 0.004
  0.408 0.001
  0.429 0.000
  0.449 0.000
  0.469 0.001
  0.490 0.000
  0.510 0.000
  0.531 0.000
  0.551 0.000
  0.571 0.000
  0.592 0.000
  0.612 0.000
  0.633 0.000
  0.653 0.000
  0.673 0.000
  0.694 0.000
  0.714 0.000
  0.735 0.000
  0.755 0.000
  0.776 0.000
  0.796 0.000
  0.816 0.000
  0.837 0.000
  0.857 0.000
  0.878 0.000
  0.898 0.000
  0.918 0.000
  0.939 0.000
  0.959 0.000
  0.980 0.000
  1.000 0.000
  /
\end{sparkline}
\\
Pop &
\begin{tabular}{@{}c@{}}%
0.035 \\%
{\smaller[2] (0.032, 0.038)}\end{tabular} &
\begin{tabular}{@{}c@{}}%
0.000 \\%
{\smaller[2] (0.000, 0.000)}\end{tabular} &
\begin{tabular}{@{}c@{}}%
0.000 \\%
{\smaller[2] (0.000, 0.001)}\end{tabular} &
\begin{tabular}{@{}c@{}}%
0.134 \\%
{\smaller[2] (0.128, 0.160)}\end{tabular} &
\begin{sparkline}{9}
\sparkdot 0.035 0.186 red
\spark
  0.000 0.992
  0.020 0.518
  0.041 0.138
  0.061 0.087
  0.082 0.081
  0.102 0.083
  0.122 0.078
  0.143 0.078
  0.163 0.091
  0.184 0.041
  0.204 0.018
  0.224 0.017
  0.245 0.012
  0.265 0.009
  0.286 0.008
  0.306 0.006
  0.327 0.003
  0.347 0.002
  0.367 0.001
  0.388 0.001
  0.408 0.000
  0.429 0.000
  0.449 0.000
  0.469 0.000
  0.490 0.000
  0.510 0.000
  0.531 0.000
  0.551 0.000
  0.571 0.000
  0.592 0.000
  0.612 0.000
  0.633 0.000
  0.653 0.000
  0.673 0.000
  0.694 0.000
  0.714 0.000
  0.735 0.000
  0.755 0.000
  0.776 0.000
  0.796 0.000
  0.816 0.000
  0.837 0.000
  0.857 0.000
  0.878 0.000
  0.898 0.000
  0.918 0.000
  0.939 0.000
  0.959 0.000
  0.980 0.000
  1.000 0.000
  /
\end{sparkline}
\\
\bottomrule
\end{tabular}

    \caption{Summary statistics of algorithm performance ($\operatorname{RBP}_{0.8}$). 
    See \S\ref{sec:study} for details.}
    \label{tab:cs:summary}
\end{table}

\begin{figure}[tbph]
    \centering
    \includegraphics[width=\textwidth]{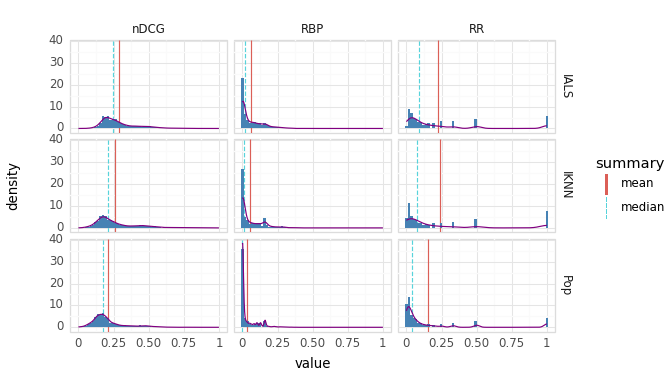}
    \caption{Distribution of per-user effectiveness scors (nDCG, $\operatorname{RBP}_{0.8}$, and reciprocal rank) as both histograms and density plots.}
    \label{fig:cs:user-score-dist}
\end{figure}

The first tool is to simply look at the distributions of performance metrics or improvements.  This is most applicable to distributions of user utility, and facilitates both inspection of a single system's distribution, comparing distributions (through parallel distribution plots), or looking at distributions of differences (by plotting the distribution of improvement in a paired evaluation).
When space permits, full histograms or kernel density plots can be shown, as in Fig.~\ref{fig:cs:user-score-dist}; it is also possible, however, to integrate distribution summaries and visualizations into the kinds of tables that are typically included in IR evaluation reports and papers.  For example, Table~\ref{tab:cs:summary} shows summary statistics for the nDCG of multiple algorithms in a recommender system evaluation; each row reports the mean score for that algorithm (as is typical practice), but also a kernel density plot of each algorithm's performance over the set of test users rendered with the \LaTeX{} \texttt{sparklines} package.
See Section~\ref{sec:study} for more detailed discussion of these results.

\begin{figure}[tbph]
    \centering
    \includegraphics[width=\textwidth]{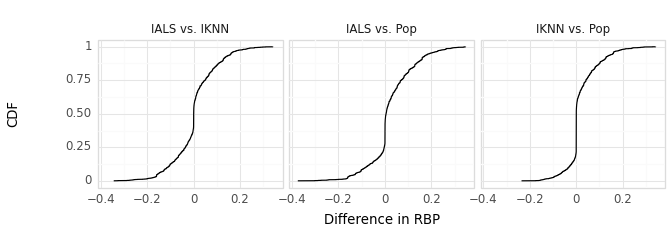}
    \caption{Empirical CDF of the the distribution of the differences $\operatorname{RBP}_{0.8}$ for the algorithms for each test user.}
    \label{fig:cs:rbp-diff-dist}
\end{figure}

\begin{figure}[t!]
    \centering
    \includegraphics[width=\columnwidth]{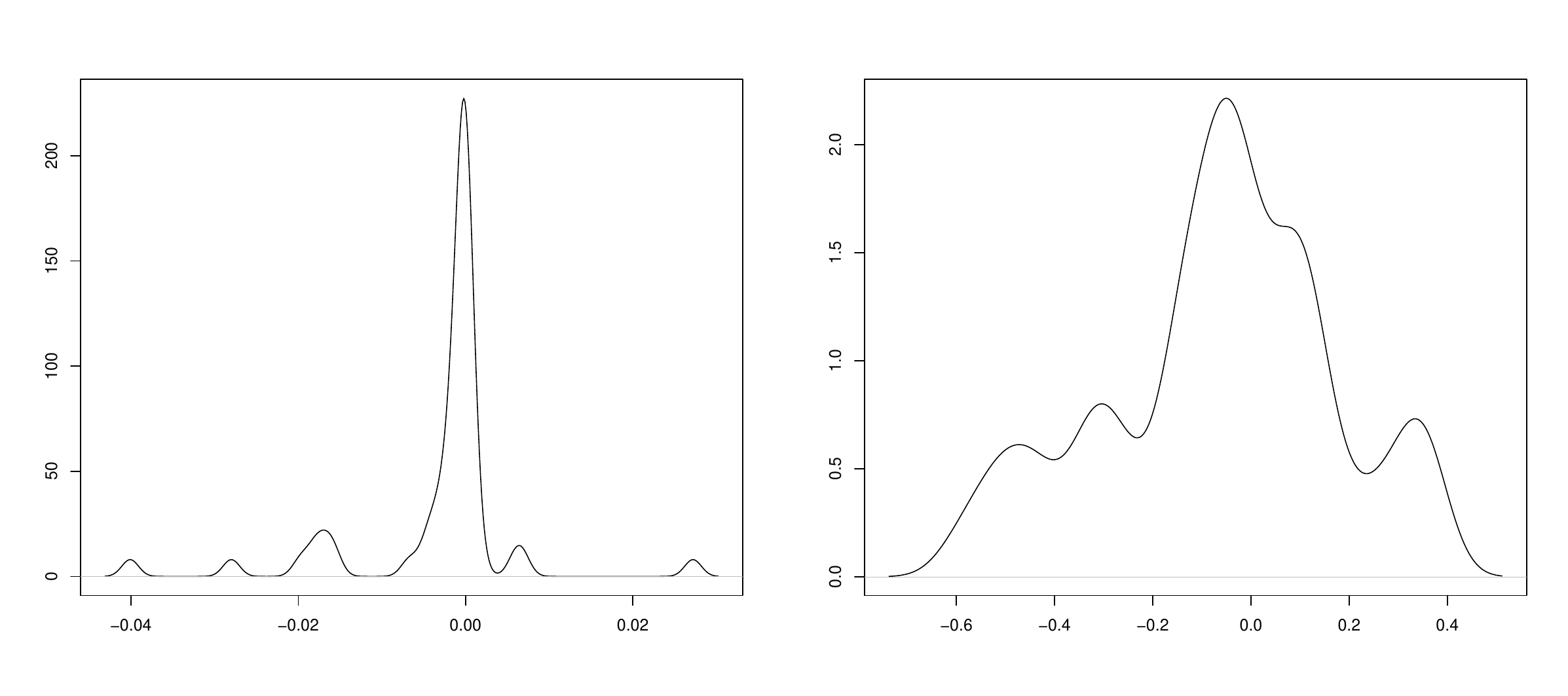}
    \caption{Two different distributions of differences in  average precision. The left ranges from -0.04 to 0.03 with the mean difference at 0.003; the right from -0.6 to 0.4 with the mean at 0.06.}
    \label{fig:distdiffs}
\end{figure}

We can also inspect the distribution of differences, in addition to quantifying it, as shown in Fig.~\ref{fig:cs:rbp-diff-dist}; this is similar to the helped-hurt histograms proposed by \citet{wangRobustRankingModels2012}.
Figure~\ref{fig:distdiffs} illustrates how distribution information can provide insights that pointwise estimates cannot.
Both plots show a kernel density of the distributions of differences between two retrieval systems submitted to the TREC 8 ad-hoc track.
The two systems in the left plot have a mean difference in mean average precision of 0.003, which is statistically significant.
The two systems in the right plot have a mean difference in MAP of 0.06; though larger, this difference is not statistically significant.
The full distributions in both cases reveal major differences: the left distribution is very constrained, with almost no variation from query to query.  Though the difference is significant, it is unlikely that end-users will detect any differences, and thus hard to ascribe any meaning to it.
The right distribution shows much more variance, in a way that is much more likely to impact end users. 
System effectiveness on some queries is as much as -0.5 lower in terms of average precision, which is sure to be impactful, yet the pointwise estimate suggests the left-hand system is better and the significance test does not convey any reason to be concerned.

\subsection{Multiple Statistics}
Another immediately-available tool is to report multiple statistics from a distribution in addition to its mean.  The median is an obvious choice; other order statistics, such as the top and bottom quartiles, deciles, or percentiles, will give further insight into where the most utility is distributed across consumers, providers, or sets of stakeholders.
As shown in Figure \ref{tab:cs:summary}, we can see this leads to different conclusions about relative overall algorithm performance.
We invite further community discussion and further research to identify generally-useful sets of statistics that will summarize distributions and enable their comparison.

Bootstrapping provides a readily-available tool for reporting confidence intervals for each of these estimates, along with differences in them (e.g. the difference in medians or the median difference between two systems), providing statistical rigor to inferences of relative system performance based on arbitrary summaries of the distribution.


\subsection{Distribution-Based Metrics}
Some recent metrics, such as expected exposure loss \citep{diazEvaluatingStochasticRankings2020}, are distributional at their heart: the metric measures the distance between the system's expected distribution of utility to the providers of documents or items and the distribution that would be expected under an ideal policy.
This is certainly not the only conceivable metric that incorporates a distribution.
Metrics for capturing the behavior of stochastic rankers, distributions over information needs, and uncertainty is a rich area for further research in IR evaluation.

There are, broadly speaking, at least four different ways we can compute distribution-based metrics:

\begin{itemize}
    \item Capturing relevant \textbf{characteristics of the distribution} itself; for a simple example, computing the inter-quartile range or the standard deviation provides a measure of the consistency of the system.
    
    \item Computing \textbf{statistics of pairs or sets of distributions} to characterize the potential impact of their differences.  For example, given two independent (non-paired) distributions of system effectiveness over user requests, we may wish to estimate the expected proportion of requests for which $S_1$ outperforms $S_2$.  We can calculate this expectation as the sum over effectiveness values $x$, the probability that $S_2$ reaches $x$ for a request times the cumulative density of requests for which $S_1$ outperforms $x$~\cite{carteretteSimulatingSimpleUser2011a}.  
When distributions are not independent, or there are sets rather than pairs, this generalizes to computations over multivariate distributions.  Carterette presented a method for comparing rankings of systems that uses distributional information in this way~\citep{carteretteRankCorrelationDistance2009}. 

    \item Comparing the \textbf{distributions from two systems}, such as the baseline system and a proposed alternative in either an online A/B trial or an offline experiment, allows us to examine differences in performance between the systems.  This can be done graphically; by comparing relevant statistics; or in some cases through distribution divergence metrics such as Jenson-Shannon and Wasserstein (although divergence between two systems is likely hard to interpret and relate to application goals).
    \item Comparing the system distribution with a \textbf{target distribution}, such as the expected exposure or utility from an omniscient ranker \citep{diazEvaluatingStochasticRankings2020} or externally-derived target distributions \citep{sapiezynskiQuantifyingImpactUser2019}.  Here divergence metrics likely make more sense, as they capture how closely the system is approximating the target.
    This is similar in spirit to the normalization of nDCG \citep{jarvelinCumulatedGainBasedEvaluation2002}, which compares the achieved utility to the ideal, but extends it to distributions and applies the concept in ways that can account for rich modeling of uncertainty.
\end{itemize}

\subsection{Confidence Measures}

When we can quantify the confidence, uncertainty, or volatility in the various metrics and scores that go into a system's outputs and evaluation (such as the confidence in feedback or annotations, or the confidence in the system's estimated relevance scores), we can feed this quantified uncertainty into a distributional evaluation to gain a more complete, end-to-end picture of its behavior that accounts for data quality and model uncertainty.
Existing and future research on estimating confidence and uncertainty across IR and machine learning pipelines will therefore be valuable for this effort.

\subsection{Monte Carlo Simulations}

Simulations of various forms have a long history in information retrieval research \cite{tagueProblemsSimulationBibliographic1980} and are increasingly applied to recommender systems as well.
There are a range of simulation applications in recommender system evaluation:
\begin{itemize}
\item Bootstrap sampling evaluation metrics to produce confidence intervals and $p$-values (simulating the sampling distribution)
\item Markov Chain Monte Carlo (MCMC) sampling for Bayesian inference over traditional evaluation metrics
\item Sampling hypothetical feedback from simulated users of a system trained on traditional data
\item Repeated model evaluation over resampled data to simulate system performance over different collections, such as sharding
\item Simulating data, allowing for estimation of the distribution of system responses over a range of data conditions
\end{itemize}

As noted in Section~\ref{sec:simulation}, the randomization in such simulation is itself a source of aleatoric uncertainty in the final results, in addition to being a useful tool for exploring uncertainty elsewhere in the information access system and its experiments. Running a simulation repeatedly, and reporting the results across multiple simulations, is a starting point for quantifying this uncertainty; for two examples, \citet{urbanoTestCollectionReliability2015} reports distributions across multiple simulation runs for test collection reliability and \citet{tianEstimatingErrorBias2020} report results over 100 runs of their simulation for measuring recommender evaluation metric error.  Sharding \citep{voorheesUsingReplicatesInformation2017} uses random partitioning or subsampling of a document collection to quantify uncertainty around the effect size of a system's performance.  \citet{chaneyHowAlgorithmicConfounding2018} ran 10 instances of their simulation, reporting averages from across the runs; results could be reported with distributions.

Monte Carlo Bayesian inference is not commonly employed in recommender systems research, but uses simulation to estimate posterior distributions of graphical models (see the next section). \citet{carteretteBayesianInferenceInformation2015} uses this technique for analyzing effectiveness scores, and \citet{ekstrandExploringAuthorGender2021} estimate distributions of author gender biases in recommender system data and results.
STAN~\citep{carpenterStanProbabilisticProgramming2017} is an effective, modern package for such inferences.

\subsection{Bayesian Modeling}
Bayesian modeling offers a framework that allows experimenters to model many different sources of uncertainty.
Using prior distributions and multi-level graphical models allows the modeling of multiple sources of uncertainty as well as the propagation of uncertainty through our reasoning about system effectiveness.
Instead of point estimates and confidence intervals, all reasoning is done on posterior distributions, which are computed from priors, observations, and explicit modeling assumptions.
By making modeling assumptions explicit, Bayesian modeling is a transparent way to conduct experimental analysis.

\begin{figure}
    \centering
    \includegraphics[width=0.45\columnwidth]{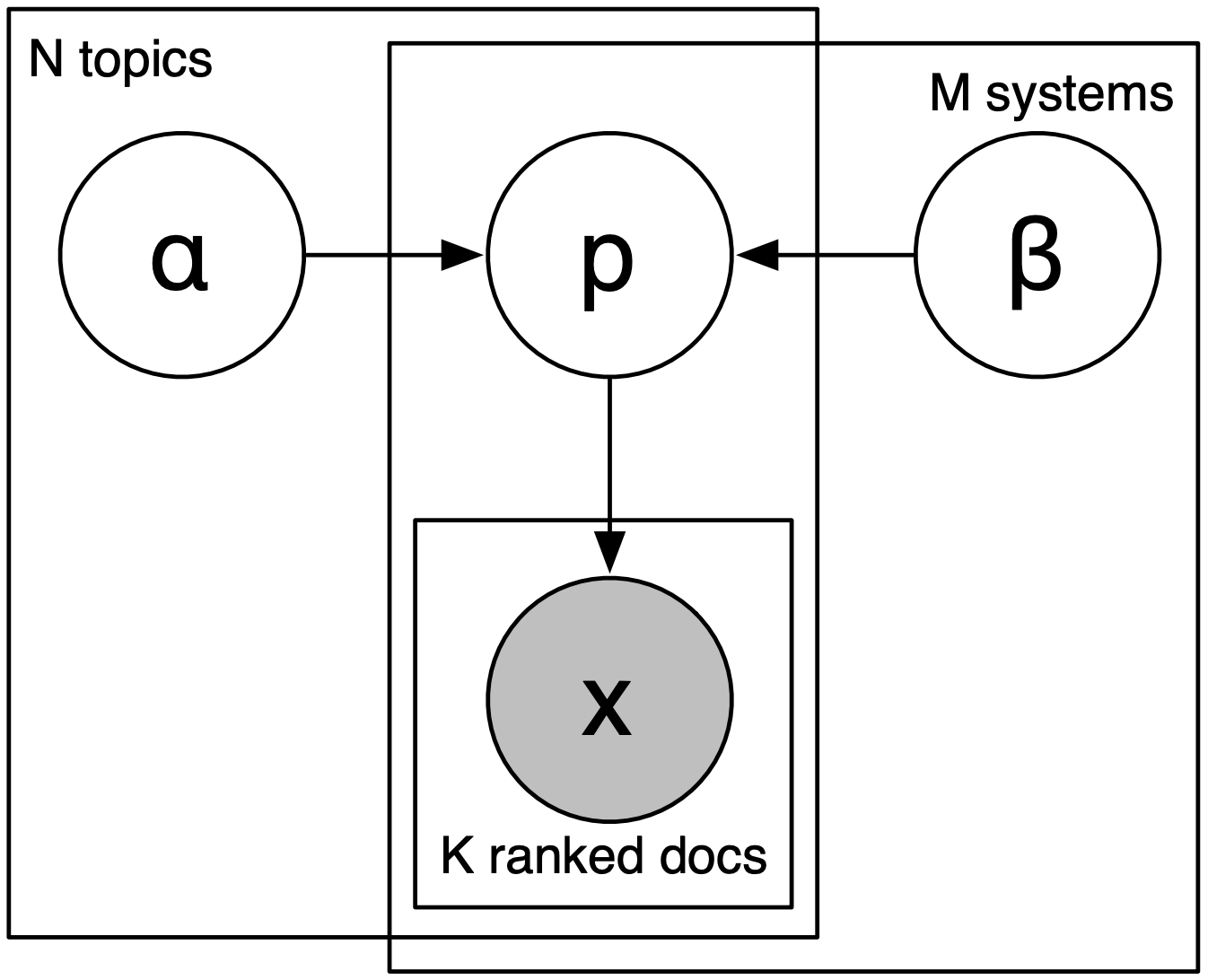}
    \caption{Graphical model of information request and system influencing the observed outcome of ranking an item at position $k$. From~\citep{carteretteBayesianInferenceInformation2015} and used with permission of author.}
    \label{fig:graphmodel}
\end{figure}

An example graphical model for search evaluation is shown in Figure~\ref{fig:graphmodel}.
In this model, the shaded node $x$ is an observation: a relevance judgment, or a click, or some other recorded indication of the usefulness of a ranked result.
This observation is modeled as the outcome of a sampling procedure from a distribution with parameter $p$; a simple case is that $x$ is a binary value and $p$ is the parameter of a Bernoulli distribution.
Then $p$ in turn is modeled as the outcome of a sampling process defined by parameters $\alpha$ and $\beta$.  
These parameters can be treated as models of topic ``hardness'' and system effectiveness respectively.
All this requires is linking the topic parameter $\alpha_j$ and the system parameter $\beta_i$ through to the item $x_{ijk}$ ranked at position $k$ by system $i$ for request $j$.

\citet{carteretteBayesianInferenceInformation2015} presents several different models of increasing complexity, incorporating additional prior distributions modeling graded relevance and user browsing behavior.
\citet{benhamBayesianInferentialRisk2020} describe a Bayesian approach to risk-sensitive retrieval, giving more weight in an evaluation to queries that under-perform relative to a baseline.




\subsection{Open-Source Research Software}

While the judgement calls informed by performance and impact distributions and application needs cannot be fully automated, there is significant room for open-source software and reusable examples to produce the kinds of metrics and reports that will support such decisions and the analyses we envision.
The case study in Section~\ref{sec:study} was prepared with LensKit, and the specific code to support it will be published with this paper.
Software such as Quarto \citep{quarto} can further help facilitate the collection of metrics and visualizations that will support an evaluation through public templates for computational documents that present common distributions and metrics.

\subsection{Likely Challenges}

We do not claim that the evaluation regime we promote will be easy or without challenges.
Rigorously evaluating recommender systems is already a complex process with significant opportunity for error; distributional evaluations will introduce further subtlety and complexity that makes it difficult to evaluate proposed improvements, or at least more difficult than comparing first-order performance metrics.
We contend, however, that this complexity is inherent to making informed decisions about whether proposed advances in recommendation algorithms will be suitable for a particular context, and for thoroughly understanding the benefits and behavior of recommender systems.
Interpreting distributions will also require sound and considered judgement as to what differences and behavior are beneficial for a particular application.
We do recommend that mean performance continue to be reported, both as one summary (among many) and for comparability with past results.
Reporting distributional analyses will provide further context for the point estimates and the decisions made in an evaluation and analysis, so that readers can better assess the appropriateness of the original decisions and their potential impact on decisions or future work that relies on the results.

There is also a computational cost to this work--- quantifying uncertainty from some sources requires re-running part or all of an experiment multiple times. 
Some repetition is necessary to ensure result reliability.
Further research will need to provide guidance about how to prioritize different uncertainty sources based on the costs of characterizing them and the likely benefit or impact on decisions that arises from that use of computational resources.

\section{Case Study}
\label{sec:study}

In this section, we present a case study that demonstrates several types of distributional analyses.
Source code for this experiment is available at \href{https://doi.org/10.5281/zenodo.8157683}{doi:10.5281/zenodo.8157683} and on GitHub\footnote{\url{https://github.com/mdekstrand/tors-distribution-eval}}.

\subsection{Experiment Description}
For our case study, we present a relatively straightforward experiment to evaluate a candidate algorithm to replace the system's existing collaborative filter.
In our scenario, the system is currently running an item-item nearest-neighbor collaborative filter in implicit-feedback mode \citep[IKNN,][]{deshpandeItembasedTopNRecommendation2004}.
The developers are proposing to replace this with an implicit-feedback matrix factorization algorithm \citep[IALS,][]{takacsApplicationsConjugateGradient2011}, and are carrying out their experiment with the LensKit toolkit \citep{lkpy}\footnote{In this experiment, we use default parameter settings from LensKit; there are open questions about how to do hyperparameter tuning under distributional evaluation (see \S\ref{sec:future-research}), but tuned algorithms will not change the process we are attempting to illustrate.}.
For reference, a basic popular-items baseline (Pop) is also included.
We evaluated each algorithm on 1500 test users with 5 held-out test ratings, generating 1000 recommendations for each.
To examine distributions over different random seeds, we ran the experiment 50 times with different data splits and initial conditions for model training (non-repeated results are reported only on the first run).

We focus on evaluating effectiveness with Rank-Biased Precision \citep[RBP,][]{rbp} with a patience parameter of $\gamma=0.8$ ($\operatorname{RBP}_{0.8}$; $r_{ui} \in \{0,1\}$ is the implicit feedback indicator variable):

\begin{align*}
    \operatorname{RBP}_\gamma(L_u)
    & = (1 - \gamma) \sum_{i=1}^N 
    r_{ui} \gamma^i
\end{align*}

We chose RBP to allow for a conceptually-consistent evaluation between both user-side utility and provider-side exposure, as the geometric browsing model in RBP is readily amenable to use in the Expected Exposure construct \citep{diazEvaluatingStochasticRankings2020}.
We chose a relatively high patience parameter to yield a decay curve that is similarly shallow to the nDCG metric used more commonly in recommender system evaluation.

The fundamental question the experiment is attempting to answer is whether or not to field IALS for an A/B test.
Similar analyses would then be done on the results of the A/B test.

\subsection{Baseline Results}

\begin{table}[tbh]
    \centering
    \begin{tabular}{lrrrrrrr}
 & $\operatorname{RBP}_{0.8}$ & $\operatorname{RBP}_{0.5}$ & HR & HR@10 & HR@20 & nDCG & MRR \\
 \toprule
IALS & \textbf{0.061} & 0.045 & \textbf{1.000} & \textbf{0.495} & \textbf{0.681} & \textbf{0.286} & 0.223 \\
IKNN & 0.057 & \textbf{0.052} & \textbf{1.000} & 0.448 & 0.594 & 0.261 & \textbf{0.237} \\
Pop & 0.035 & 0.030 & 0.996 & 0.302 & 0.452 & 0.211 & 0.155 \\
\midrule
$p$ (IALS-IKNN) & 0.071 & 0.030 & NA & 0.005 & $<0.001$ & $<0.001$ & 0.172
\\
\bottomrule
    \end{tabular}
    \caption{Point estimates of system performance with $p$-values from paired $t$ tests between IALS and IKNN.}
    \label{tab:cs:point-estimates}
\end{table}

Table~\ref{tab:cs:point-estimates} shows the basic point-estimate evaluation results we would obtain in a typical evaluation, showing $\operatorname{RBP}_{0.8}$ along with several other evaluation metrics.
We see in these results that IALS outperforms IKNN on our primary metric, and all metrics except for untruncated HR, MRR, and RBP with $\gamma=0.5$; many metrics yield a statistically significant difference ($p < 0.01$).
A typical evaluation focused on the selected metric, or on nDCG or hit rate on reasonably short lists, would conclude that IALS should advance to A/B trials; the high $p$-value under the target metric gives pause, but the developers may choose to try the system online anyway.

\subsection{Basic Distributional Reporting}

Table~\ref{tab:cs:summary} shows the results on $\operatorname{RBP}_{0.8}$ for the three algorithms with more complete distributional statistics: mean, median, percentiles, and bootstrapped confidence intervals for each, along with a KDE plot of the distribution of algorithm performance over users.
This shows that not only does IALS outperform IKNN in mean performance, but its median and max performance are also better.
Fig.~\ref{fig:cs:user-score-dist} shows more detailed distributions of per-user data for three of the metrics.

\subsection{Distribution of Differences}

Fig.~\ref{fig:cs:rbp-diff-dist} shows the ``distribution of differences'': the
empirical cumulative distribution of the per-user differences in
$\operatorname{RBP}_{0.8}$ between pairs of algorithms. The median difference
between IALS and IKNN is $3.7 \times 10^{-5}$, so IALS is better than IKNN for a
majority of users.  Approximately 30\% of users do have worse recommendations
under the new algorithm, however.

\subsection{User Subgroup Distributions}

\begin{figure}[tbph]
    \centering
    \begin{minipage}{0.48\textwidth}
        \includegraphics[width=\textwidth]{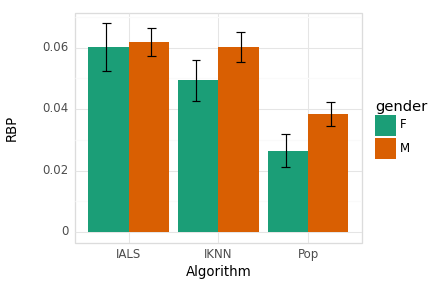}
        \subcaption{Mean performance with 95\% CIs.}
        \label{fig:cs:rbp-gender:perf}
    \end{minipage}
    \begin{minipage}{0.48\textwidth}
        \includegraphics[width=\textwidth]{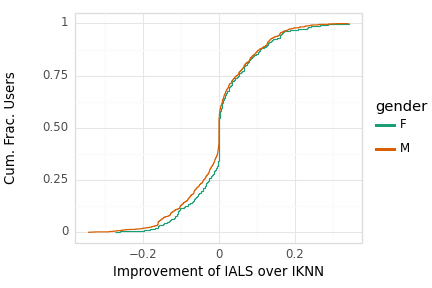}
        \subcaption{Distribution of per-user change in performance.}
        \label{fig:cs:rbp-gender:change}
    \end{minipage}
    \caption{Mean effectiveness ($\operatorname{RBP}_{0.8}$) disaggregated by user gender. We see in (a) that most of IALS's improvement in the top-line evaluation score comes improvements to recommendations for female users, who had noticeable lower-quality recommendations from IKNN and Pop.  This is consistent with the distribution of per-user improvements in (b); since the female curve is slightly to the right of the male curve, we can see that female users have slightly more than male users, and this is fairly consistent instead of coming through improving things for a few women while harming them for others.}
    \label{fig:cs:rbp-gender}
\end{figure}

Table~\ref{fig:cs:rbp-gender:perf} shows the effectiveness ($\operatorname{RBP}_{0.8}$) disaggregated by user gender. It shows that the current system (IKNN) has a notable gap in gender performance, which is closed by the IALS algorithm; further, most of IALS's improvement in mean performance comes from improving performance for female users, and a $t$-test for the improvement on female users yields $p=0.0134$.

\subsection{Distribution over Uncertain Parameters}

\begin{figure}[tbph]
    \centering
    \begin{minipage}{0.48\textwidth}
        \includegraphics[width=\textwidth]{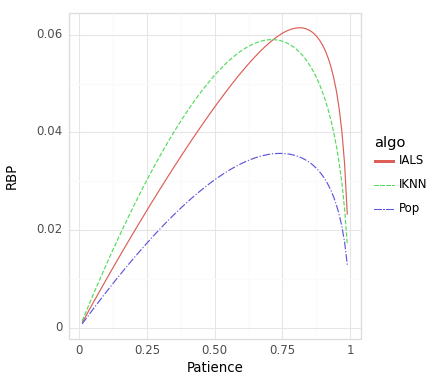}
        \subcaption{Mean RBP as a function of patience.}
        \label{fig:cs:rbp-uncertain:func}
    \end{minipage}
    \begin{minipage}{0.48\textwidth}
        \includegraphics[width=\textwidth]{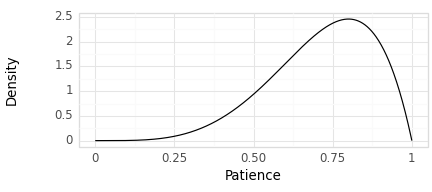}
        \includegraphics[width=\textwidth]{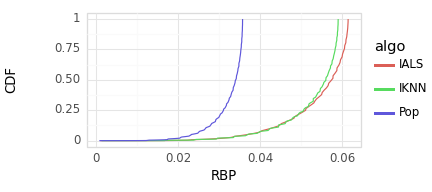}
        \subcaption{Patience prior \& RBP posterior.}
        \label{fig:cs:rbp-uncertain:bayes}
    \end{minipage}
    \caption{Mean RBP for algorithms with different patience parameters. (\subref{fig:cs:rbp-uncertain:func}) shows how the mean RBP changes in response to patience; we can see that IALS performs better with large patience models, but IKNN remains better when the patience value is decreased. \subref{fig:cs:rbp-uncertain:bayes} shows a Bayesian analysis in which we model prior knowledge of the browsing model as a distribution over patience values (top), and the resulting posterior distributions of RBP (bottom); we can see that IALS has more probability mass on higher values, suggesting a posterior belief in favor of IALS. This prior is purely for illustrative purposes.}
    \label{fig:cs:rbp-uncertain}
\end{figure}

The distributions we have presented so far are distributions over samples, either users or subgroups; this is a form of aleatoric uncertainty, in that the arrival of users at the system is effectively a random process (or can be treated as such).
Distributions can also engage with epistemic uncertainty, however.
Fig.~\ref{fig:cs:rbp-uncertain} shows these results.
In Fig.~\ref{fig:cs:rbp-uncertain:func}, we see how the effectiveness scores change as the patience parameter changes; IALS outperforms IKNN when $\gamma$ exceeds approximately 0.72.
Not all values of $\gamma$ are equally likely, however; we can also represent our epistemic uncertainty as a prior distribution; for illustration we have chosen a Beta distribution whose mode is the original value of 0.8 ($\operatorname{Beta}(5, 2)$.
Fig.~\ref{fig:cs:rbp-uncertain:bayes} shows this prior along with the CDFs of the effectiveness metrics arising from this prior, showing that IALS performs at least as well as KNN, if not better, across the bulk of the probability mass (the most likely values for $\gamma$).

\begin{figure}[tbph]
    \centering
    \includegraphics[width=\textwidth]{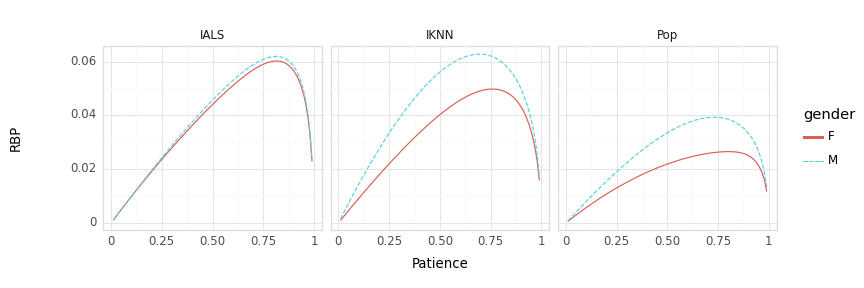}
    
    \caption{RBP as a function of patience, as in Figure~\ref{fig:cs:rbp-uncertain:func}, disaggregated by gender.  We see that female users' recommendation effectiveness is improved from IKNN to IALS across a range of patience parameters, indicating that the closing of the gender gap is robust across browsing model parameter choices.}
    \label{fig:cs:rbp-uncertain:gender}
\end{figure}

We can further disaggregate by users (or other stakeholders). Fig.~\ref{fig:cs:rbp-uncertain:gender} illustrates Fig.~\ref{fig:cs:rbp-uncertain:func} disaggregated by user gender, showing that IALS's closing of the gender gap in system effectiveness holds across browsing model parameters.

\subsection{Item Distribution}

\begin{figure}
    \centering
    \begin{minipage}{0.48\textwidth}
    \centering
    \includegraphics[width=\textwidth]{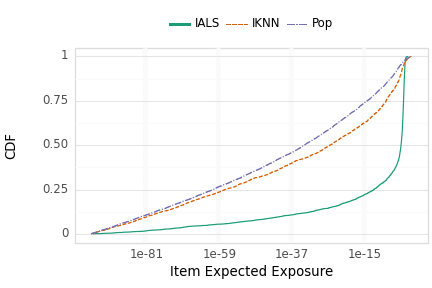}
    \subcaption{CDF of item exposure.}
    \label{fig:cs:item-exp:cdf}
    \end{minipage}
    \begin{minipage}{0.48\textwidth}
    \centering
    \includegraphics[width=\textwidth]{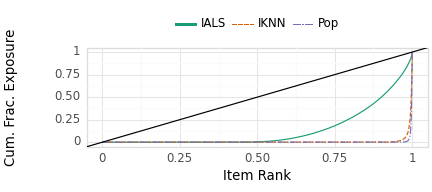}
    \subcaption{Lorenz curve of item exposure.}
    \label{fig:cs:item-exp:lorenz}
        {\small
        \begin{tabular}{lrrr}
            \toprule
            {} &      Gini &        $L_2$ &        KL \\
            \midrule
            IALS &  0.732345 &  0.000329 &  0.424431 \\
            IKNN &  0.988025 &  0.025326 &  1.993307 \\
            Pop  &  0.995073 &  0.051245 &  2.398029 \\
            \bottomrule
            \end{tabular}}
        \subcaption{Item exposure distribution statistics.}
        \label{fig:cs:item-exp:stats}
    \end{minipage}
    \caption{Distribution of expected exposure of individual items, displayed as both an empirical CDF and as a Lorenz curve (used for computing Gini coefficients), along with statistics of the distribution (Gini) and comparison of the item exposure distribution to that of an ideal ranking policy ($L_2$ and KL).
    We can see that IALS is distributing much more exposure to a larger set of distinct items.}
    \label{fig:cs:item-exp}
\end{figure}

We also consider the distribution of benefit to another stakeholder class, the items themselves (which can be easily extended to the providers of these items).
Expected Exposure~\citep{diazEvaluatingStochasticRankings2020} provides a way to measure the exposure that accrues to each item, using the same browsing model as used in RBP.
Fig.~\ref{fig:cs:item-exp:cdf} shows the distribution of per-item exposure across the test users for each system.
We see that both Pop and IKNN have many items with relatively low exposure; IALS has many more items with relatively high exposure, indicating that it is distributing exposure considerably more equally between items and demonstrates less popularity bias.
This can be seen in alternate form from the Lorenz curves in Fig.~\ref{fig:cs:item-exp:lorenz} and the Gini coefficients in Fig.~\ref{fig:cs:item-exp:stats}, where IALS is substantially closer to equality than either IKNN or Pop.

\citet{diazEvaluatingStochasticRankings2020} also compare a system's exposure to that of an ideal target policy that distributes expected exposure equally across relevant items for a particular user, which facilitates a fairness goal that an item or provider's exposure should be commensurate with their relevance or utility.
A plot of the distribution of individual item comparisons to the results of this policy was not very instructive, but Fig.~\ref{fig:cs:item-exp:stats} shows the results of comparing each algorithm's exposure distribution to that of the ideal policy with both the $L_2$ metric used by \citeauthor{diazEvaluatingStochasticRankings2020} and K-L divergence, showing that IALS not only distributes exposure more equally across items, it distributes it more equally across relevant items.

\subsection{Item Subgroups}

\begin{figure}
    \begin{minipage}{0.48\textwidth}
    \centering
    \includegraphics[width=\textwidth]{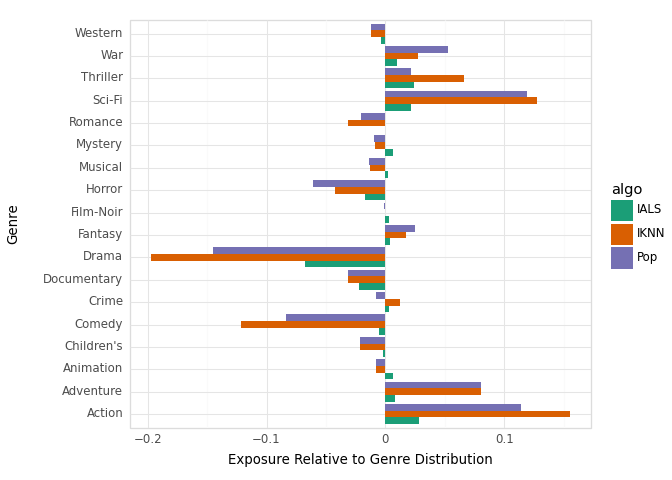}
    \subcaption{Item exposure by genre relative to genre prevalence.}
    \label{fig:cs:genre-exp:prev}
    \end{minipage}
    \begin{minipage}{0.48\textwidth}
    \centering
    \includegraphics[width=\textwidth]{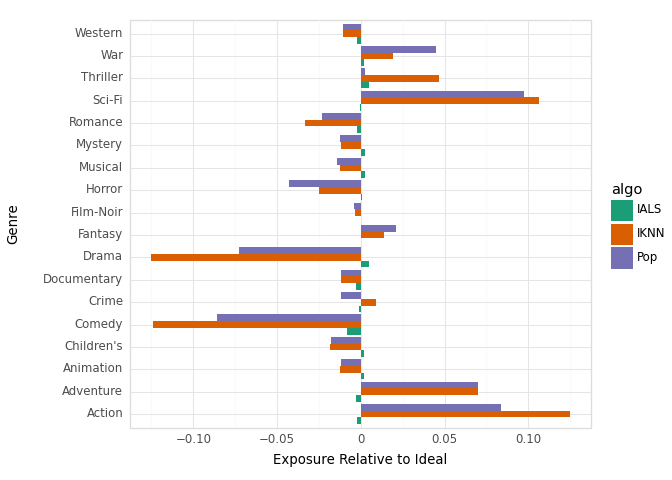}
    \subcaption{Item exposure by genre relative to ideal.}
    \label{fig:cs:genre-exp:ideal}
    \end{minipage}
    \caption{Item exposure by genre.  These plots compare with two reference points: \subref{fig:cs:genre-exp:prev} compares the distribution of genre exposure to the prevalence of that genre in the data set (how many movies have the genre, fractionalized for movies with multiple genres), and \subref{fig:cs:genre-exp:ideal} compares it to the exposure for movies of that genre under an ideal policy.}
    \label{fig:cs:genre-exp}
\end{figure}

As an example of an item subgroup analysis, we have aggregated exposure by movie genre as recorded in the MovieLens data set (using fractional membership to handle movies with multiple genres).
Fig.~\ref{fig:cs:genre-exp} shows the distribution of total exposure per genre, relative to two reference points: the distribution of genres in the corpus of movies, and the distribution of exposure to genres under an ideal ranking policy.
IALS does a better job of matching both distributions, as can be seen by the bars closer to 0, and this is confirmed by both $L_2$ (0.0002 for IALS vs. 0.0684 for IKNN, with respect to ideal) and K-L divergence (0.0017 vs. 0.3563).

\subsection{Repeated Evaluation}

\begin{figure}[tbph]
    \begin{minipage}{0.48\textwidth}
    \centering
    \includegraphics[width=\textwidth]{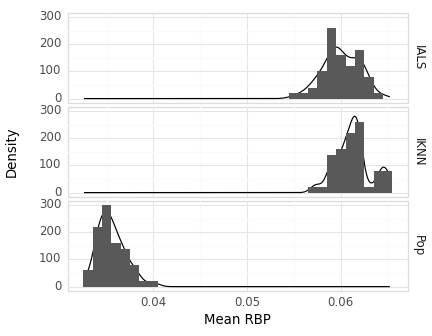}
    \subcaption{Distribution of mean $\operatorname{RBP}_{0.8}$ over repetitions.}
    \label{fig:cs:repeat:mean}
    \end{minipage}
    \begin{minipage}{0.48\textwidth}
    \centering
    \includegraphics[width=\textwidth]{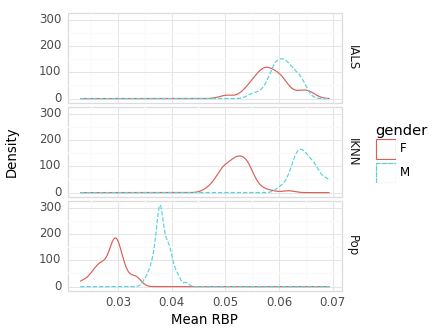}
    \subcaption{Distribution of means disaggregated by gender.}
    \label{fig:cs:repeat:gender}
    \end{minipage}
    \caption{Distribution of overall performance ($\operatorname{RBP}_{0.8}$) across multiple repetitions of the evaluation.}
    \label{fig:cs:repeat}
\end{figure}

The final distributions we show are over repeated runs of the evaluation.
Fig.~\ref{fig:cs:repeat} shows the mean $\operatorname{RBP}_{0.8}$ across 50 repetitions of the evaluation with different test set samples and initial values for model training.
This indicates that the improvement in performance as measured by $\operatorname{RBP}_{0.8}$ is not stable, consistent with the lack of statistical significance; the closing of the gender gap and improvement for female users do look to be stable across repetitions, however, so we may still wish to field IALS for trial; in other seeds, performance for male users may be slightly degraded, however.
Similar plots can be drawn for distributions of differences, exposure statistics, and other measures.

\subsection{Summary}
\label{sec:study:summary}

In our example experiment and decision process, most evaluation metrics agreed that the IALS algorithm outperforms the IKNN baseline, with the exception of two metrics that emphasize the top of the recommendation list to a much greater degree ($\operatorname{MRR}$ and $\operatorname{RBP}_{0.5}$).
However, our distributional analysis yielded significant insights into \textit{why} IALS performed better, and provide guidance that support a decision to field it:

\begin{itemize}
    \item Female users, who had significantly lower-quality recommendations under IKNN, see the most improvement under IALS.
    \item This improvement does not come with degradation in quality, on average, for male users.
    \item IALS provides substantially more equitable allocation of exposure, both to individual movies and movie genres, than either IKNN or Pop.
    \item The closing of the gender gap in recommendation performance is robust to changes in the RBP patience parameter.
    \item When accounting for the a priori plausibility of different patience values, the posterior distribution of performance favors IALS.
    \item Overall relative performance is not stable across repetitions, but the reduction in the gender gap in recommendation effectiveness is.
\end{itemize}

The natural interpretation for the discrepancy in relative performance in top-level point estimates for different metrics is that IALS puts more relevant items in reasonably high positions in the ranking, while IKNN may be better at putting one relevant item very high in the ranking.
The details of the target application will determine which is more important, but assuming that placing multiple relevant items in recommendation lists is desirable, the distributional analysis provides multi-faceted evidence evidence that IALS may be a better choice than IKNN (as currently configured), even though the difference in the point estimates of the primary evaluation metric was not statistically significant at $\alpha=0.05$.
Further, if we had only looked at the results of a significance test for the primary metric and rejected the proposed algorithm, we would have \emph{missed} an opportunity to deliver significant improvements both in performance for female users and equity of exposure without --- on average --- reducing effectiveness for male users.
Care is needed to ensure that this exercise does not devolve to fishing or $p$-hacking, but we believe that providing such observations in the context of a thorough distributional account of system performance (as opposed to cherry-picking a few examples) will provide transparency and context to readers and decision-makers to help them decide how highly to weight the observed subgroup improvements.
In our example, the improvements accrue along socially-salient directions (user gender and item popularity), and there are multiple different perspectives that corroborate a possible conclusion to field-trial IALS.

This analysis also yielded some tension between perspectives: IALS provided significant improvements for underserved users and item providers, without statistically disadvantaging the users who are already getting good recommendations, but its overall potential performance improvement was not stable.
Experiments require careful analysis in the context of the application, business goals, and stakeholder needs in order to assess and weigh the impact on various parties.
Distributional analysis provides a robust starting point from which to carry out that balancing process by identifying and quantifying the impacts in different directions.
It can also help with identifying where further refinement is needed --- for example, since stability of improvement is the biggest problem with an IALS conclusion, would adjusting the training settings (e.g. increasing epochs) improve its stability?

Finally, this analysis is for illustrative purposes.
There are definitely more and different distributions that could be computed and displayed.
The set that is most useful is likely to differ between applications, and we invite extensive research and community discussion about how to decide which distributions to prioritize or emphasize in any particular application.
However, it demonstrates that we can gain much deeper insight into algorithm performance and differences in algorithm performance that can inform more robust decision-making and research conclusions.

One substantial challenge facing distributional analysis is that it requires significant space to report many various distributions.
This is not a problem for internal evaluation reports, as with good document design they can be quite long and technologies such as Quarto\footnote{\url{https://quarto.org}} can facilitate the creation of standard templates for such reports that integrate into evaluation workflows.
For published research, adopting distributional evaluation will likely require greater use of appendices or supplementary material: authors can provide the main results in the paper itself, and provide a more comprehensive report of the distributional evaluation as a supplementary document in both the review process (when facilitated by the paper submission system) and final publication.

\section{Implications and Next Steps}

Adopting distributional thinking for evaluating and understanding recommender systems has implications across the range of activities associated with recommender system research, development, and deployment.

\subsection{Current Practice}

For current recommender system evaluation practice, adoption of our argument has (at least) the following implications:

\begin{itemize}
    \item We must consider the \textbf{marginal distribution} of utility within each stakeholder class.
    Does a system produce comparable utility for many of its users or subjects, or is there a substantial tail of under-served users, content producers, or other stakeholders? Does most of the benefit accrue to a few people or organizations?
    
    \item We must consider \textbf{alternate statistics} and \textbf{multiple statistics} that capture important aspects of utility distributions that are obscured in simple means; as shown in Table~\ref{tab:cs:summary}, it is quite possible for a system with higher mean performance to actually perform worse for a majority of users.
    
    \item We must consider the \textbf{distribution of subgroup aggregations} of utility.
    Does a system systematically under-serve particular minority groups of users, or content creators working in certain genres?
    There is a significant difference in the social impact of a high-variance system whose low utility is randomly distributed vs. one whose low utility disproportionately affects users already poorly-served by information retrieval systems (or other technology).
    
    \item We must consider the \textbf{distribution of differences} in utility or performance, at least when paired observations are available.
    When we have access to the utility that systems $A$ and $B$ provide to the same stakeholders, how is the improvement (or loss) in utility distributed? Do a few stakeholders experience significantly better outcomes than before while most have comparable, or even worse, outcomes?
    Do the improvements primarily accrue to those the system already serves well, or to participants currently experiencing relatively poor utility?
    How are utility gains or losses distributed with respect to salient subgroups of different stakeholder classes?
    
    \item We must consider the \textbf{difference in distributions} in utility or performance, particularly when paired observations are not available.
    Sometimes, this involves comparing the utility distributions of two systems: for example, in a within-subjects $A/B$ test, how do the distributions of the two systems compare?  Does one provide more consistent performance, or do fewer participants experience abnormally bad performance?
    Two systems may have the same mean utility, but one has more consistent performance and therefore results in fewer failed experiences.
    In other cases, we may compare a system's distribution to an ideal or target distribution, as in expected exposure \citep{diazEvaluatingStochasticRankings2020}: how closely does the system match the distribution of utility that would be expected from a perfect oracle?
    This applies both to individual-level distributions and subgroup-level distributions.
    
    \item We must consider the \textbf{distribution of impact over repeated runs}, rather than looking only at single-shot rankings.  Users rarely experience a system as a single static result; while there is value in stability \citep{adomaviciusStabilityRecommendationAlgorithms2012}, temporal diversity can provide users with more varied experiences \citep{lathiaTemporalDiversityRecommender2010}, and changing rankings over time is vital to providing fair exposure to different content providers in the presence of position bias \citep{biegaEquityAttentionAmortizing2018,diazEvaluatingStochasticRankings2020}.

    \item In production systems, these distributions should be \textbf{monitored over time}. Even if the system's overall performance in terms of aggregate utility or user satisfaction metrics does not degrade, the distribution of the system's effects may not be stable.
\end{itemize}

There is also a question of how existing or future metrics connect with distributional analysis. Any metric that computes results at a per-sample level can be analyzed with sample or subgroup distributions.
Parameters for any metric can also be modeled with distributions representing their uncertainty.
Modular metrics, such as RBP and Expected Exposure Loss, facilitate measurements that are consistent across multiple stakeholders (e.g. by using the same position-weighting model).

Examining distributions, through graphical comparison and metrics that capture more aspects of effectiveness distributions than a simple mean (such as distribution differences and carefully-chosen order statistics), will help IR and recommender system evaluation move beyond treating users, producers, and other stakeholders as interchangeable.
As can be seen in Table~\ref{tab:cs:summary}, this analysis can significantly complicate the task of determining which system is ``best'', but it is a vital part of ensuring that system improvements do not leave some participants behind or treat their experience as expendable for the sake of an overall aggregate, and lays the basis for examining where different users may actually need different system designs in order to have quality access to information.

We would also like to note that, while we envision experiments quantifying uncertainty throughout the entire data generating and experimental processes in final evaluations, we do not believe completely describing uncertainty is necessary to begin examining the distributions currently available; this examination will provide  richer insight into system behavior, performance, and impact than current standard practice, and can be incrementally expanded to account for more sources of uncertainty.

\subsection{Future Research}
\label{sec:future-research}

Distributional thinking is not simply a matter of applying known or widely-understood techniques to the results of an evaluation. Further research is needed to understand how best to report and summarize distributions in ways that actionably capture the range of a system's effects on its various users.
Several areas of research seem immediately apparent, including:

\begin{itemize}
    \item What metrics and summary statistics usefully capture the distributional effects of a system within a stakeholder class or across stakeholder classes?  There are several promising directions here, including the Expected Exposure construct \citep{diazEvaluatingStochasticRankings2020} and its multi-sided extension \citep{wuJointMultisidedExposure2022} along with positive-sum aggregation of utility across user subgroups \citep{wangUserFairnessItem2021}.
    
    \item How do we quantify and accurately characterize the uncertainty and variance that arises at different stages of the recommendation and user interaction processes?  \citet{carteretteBayesianInferenceInformation2015} discusses how to incorporate such uncertainty into an evaluation paradigm, and there is significant research on the impact of specific types of biases such as popularity bias \citep{canamaresShouldFollowCrowd2018,canamaresProbabilisticReformulationMemoryBased2017,ekstrandSturgeonCoolKids2017} and the missing-not-at-random nature of recommender systems data \citep{steckTrainingTestingRecommender2010,marlinCollaborativeFilteringMissing2007,yangUnbiasedOfflineRecommender2018}, but much work remains to characterize these and other effects into computationally-useful representations of uncertainty that can be incorporated into the recommender system evaluation process.

    \item How do we provide comparable measurements between different stakeholder groups? For example, while we used the same position weighting model for user- and item-side utility, RBP and Expected Exposure Loss are not directly comparable, so it is difficult to evaluate potential tradeoffs between users and items should they arise.

    \item What guidance can be provided for making principled, distributionally-informed decisions in various application and business contexts?  How can business, social, regulatory, and other objectives and requirements be translated into summary statistics and decision processes?
    We submit that thorough reporting of distributions will be an important enabling mechanism for such analyses, but the precise mechanisms need significant further research.

    \item How does distributional thinking interact with other experimental and deployment concerns? For example, do some data splitting strategies enable more effective analyses than others? Are multiple strategies in the same experiment needed in order to provide a thorough accounting of system behavior? Stratified sampling may be useful for characterizing the system behavior for some user or item groups, but further research is needed to understand precisely how.
    
    Hyperparameter tuning is also a significant challenge that needs additional research, as automated processes typically depend on a single statistic that can be optimized.  Are there additional statistics that can capture enough particular parameters of interest to perform tuning? Drawing from multi-objective optimization, can we automate distributional optimizations of useful forms, and if so how?

    \item How do we effectively and rigorously employ simulation in recommender systems evaluation?  There is currently a body of ongoing work on simulation for recommender systems and related research \citep{balogReport1stSimulation2021,ekstrandSimuRecWorkshopSynthetic2021,mcinerneyAccordionTrainableSimulator2021,rohdeRecoGymReinforcementLearning2018}, some of which is explicitly aimed at quantifying uncertainty \citep{mladenovRecSimNGPrincipled2021}.
    The vision we propose will have a symbiotic relationship with this line of research: such simulations, as we have noted in Section~\ref{sec:simulation}, provide a source of uncertainty over which we may want to analyze the distribution of system behavior, and the metrics and techniques developed to enable rigorous and thorough evaluation that accounts for distributions of effects and benefits will be valuable for reporting the results of such simulations.
\end{itemize}

\subsection{Paradigms and Culture}

Beyond the direct practical implications on how evaluations are carried out, and the research necessary to fully realize the vision we propose, distributional thinking has further implications for \textit{how} research and practice is approached, and the evaluation culture and community expectations for recommender systems research.
These include:

\begin{itemize}
    \item Expecting evaluations to go beyond improving the mean of an established performance metric --- researchers can provide, and reviewers can expect, more thorough accounting of the distribution of performance and performance improvements, and scrutinize results that improve the mean (or another single pointwise estimate) but do so at the expense of vulnerable or otherwise important stakeholder subgroups.

    \item Systematically looking for improved subgroup performance; existing research sometimes targets or highlights performance improvements for particular sets of users or items, either to supplement or in the absence of overall performance improvements.
    Robust distributional thinking will provide a conceptual framework for identifying, highlighting, and assessing such improvements, and we hope the analysis in Section~\ref{sec:sources} will aid in that endeavor.
    As noted in Section~\ref{sec:study:summary}, experimenters must be careful to avoid fishing or cherry-picking, but providing a thorough distributional analysis will provide context for interpreting their claims and for authors to make an argument for \textit{why} particular subgroups are relevant to consider beyond the existence of improved performance (for example, by closing the clear gender gap in performance in our case study).

    \item Shifting away from leaderboard-style research focused on improving SOTA (state-of-the-art) on established tasks in favor of scientifically and comprehensively \emph{understanding} the behavior and distribution of effects of a system, particularly in scientific publication.
\end{itemize}

On this last point, we acknowledge and appreciate the great benefit that leaderboards such as the RecSys Challenge bring to the field, particularly in giving research groups an opportunity to test their skills and new groups a platform for demonstrating their abilities.
They are valuable on-ramps to the recommender systems community.
What we hope to work with the community to promote is (1) scaffolds to help teams take the steps to move beyond optimizing a challenge's OEC (overall evaluation criterion) to thorough reporting, and (2) challenges and competitions that promote multi-perspective and distributional evaluation of systems.
Two useful steps in this direction are the incorporation of a fairness objective in the 2021 RecSys Challenge, and the multi-metric ``rounded'' evaluation used in the EvalRS AnalytiCup at CIKM 2022 \citep{tagliabueEvalRSRoundedEvaluation2022}, as well as TREC's focus on benchmarks as a means of understanding tasks and the behavior and capabilities of proposed systems~\citep{voorheesCoopetitionIRResearch2021, soboroffDatasetsWereNot2021}.

\section{Conclusion}

In conclusion, we argue that the future of recommender evaluation needs to move beyond point estimates, particularly means, of system performance or utility and attend to the \emph{distribution} of that utility --- and other system impacts --- across and within different groups of stakeholders.
This argument also applies beyond recommender systems, as all information access systems, including search engines and information filters, have similar concerns and will benefit from distributional evaluation.

Information access should be beneficial and its benefits should be equitably distributed, and attending to the distributions of effects will help make that a reality.

\begin{acks}
We thank the many collaborators and colleagues with whom we have discussed the ideas in this paper over the years.
Michael Ekstrand's contributions to this work were supported by the National Science Foundation under grant IIS 17-51278.
\end{acks}

\bibliographystyle{ACM-Reference-Format}
\bibliography{zotero.bib}

\end{document}